\RequirePackage[mathlines]{lineno}

\documentclass[prd,twocolumn,nofootinbib,showpacs,amsmath,amssymb]{revtex4-2}
\usepackage{overpic,graphicx}
\usepackage{dcolumn}
\usepackage{bm}
\usepackage{float}
\usepackage{rotating}
\usepackage{subfigure}
\usepackage{color}
\usepackage{soul}
\usepackage[bookmarksnumbered, pdfstartview=FitH,colorlinks,urlcolor=blue, citecolor=blue,linkcolor=blue,] {hyperref}
\usepackage{epstopdf}
\usepackage{multirow}
\usepackage{amsmath,esint}
\usepackage{makecell}
\usepackage{cuted}

\let\oldequation\equation
\let\oldendequation\endequation

\renewenvironment{equation}
{\linenomathNonumbers\oldequation}
{\oldendequation}

\begin{document}


\title{\bf \boldmath
Search for $\eta_c (2S)\to\pi^{+}\pi^{-}\eta_{c}$ and $\eta_c (2S)\to\pi^{+}\pi^{-}K^0_S K^{\pm}\pi^{\mp}$ decays}

\author{M.~Ablikim$^{1}$, M.~N.~Achasov$^{4,c}$, P.~Adlarson$^{75}$, O.~Afedulidis$^{3}$, X.~C.~Ai$^{80}$, R.~Aliberti$^{35}$, A.~Amoroso$^{74A,74C}$, Q.~An$^{71,58,a}$, Y.~Bai$^{57}$, O.~Bakina$^{36}$
% ...
\\
\vspace{0.2cm}
(BESIII Collaboration)\\
\vspace{0.2cm} {\it
$^{1}$ Institute of High Energy Physics, Beijing 100049, People's Republic of China\\
% ...
$^{3}$ Bochum  Ruhr-University, D-44780 Bochum, Germany\\
$^{4}$ Budker Institute of Nuclear Physics SB RAS (BINP), Novosibirsk 630090, Russia\\
% ...
$^{35}$ Johannes Gutenberg University of Mainz, Johann-Joachim-Becher-Weg 45, D-55099 Mainz, Germany\\
$^{36}$ Joint Institute for Nuclear Research, 141980 Dubna, Moscow region, Russia\\
% ...
$^{57}$ Southeast University, Nanjing 211100, People's Republic of China\\
$^{58}$ State Key Laboratory of Particle Detection and Electronics, Beijing 100049, Hefei 230026, People's Republic of China\\
% ...
$^{71}$ University of Science and Technology of China, Hefei 230026, People's Republic of China\\
% ...
$^{74}$ University of Turin and INFN, (A)University of Turin, I-10125, Turin, Italy; (B)University of Eastern Piedmont, I-15121, Alessandria, Italy; (C)INFN, I-10125, Turin, Italy\\
$^{75}$ Uppsala University, Box 516, SE-75120 Uppsala, Sweden\\
% ...
$^{80}$ Zhengzhou University, Zhengzhou 450001, People's Republic of China\\
\vspace{0.2cm}
$^{a}$ Deceased\\
% ...
$^{c}$ Also at the Novosibirsk State University, Novosibirsk, 630090, Russia\\
% ...
}
}

\begin{abstract}{
Based on $(27.12\pm 0.14)\times 10^{8}$ $\psi(2S)$ events collected with the BESIII detector, we search for the decay $\eta_c (2S) \rightarrow \pi^{+} \pi^{-} \eta_c$ with $\eta_c\rightarrow K_S^0 K^{\pm}\pi^{\mp}$ and $\eta_c\rightarrow K^{+}K^{-}\pi^{0}$. No significant signal is observed, and the upper limit on the product branching fraction $\mathcal{B}(\psi(2S)\rightarrow \gamma \eta_{c}(2S))\times\mathcal{B}$($\eta_c (2S) \rightarrow \pi^{+} \pi^{-} \eta_c$) is determined to be $2.21\times10^{-5}$ at the 90\% confidence level. In addition, the analysis of the process $\psi(2S)\to\gamma \eta_{c}(2S), \eta_{c}(2S)\rightarrow \pi^{+}\pi^{-}K^{0}_{S}K^{\pm}\pi^{\mp}$ gives a clear $\eta_c(2S)$ signal with a statistical significance of $10\sigma$ for the first time,
and the branching fraction $\mathcal{B}(\eta_{c}(2S)\rightarrow \pi^{+}\pi^{-}K^{0}_{S}K^{\pm}\pi^{\mp})$ is determined to be ($1.33 \pm 0.11 \pm 0.4 \pm 0.95 $)$\times 10^{-2}$, where the first uncertainty is statistical, the second is systematic, and the third uncertainty is due to the quoted $\mathcal{B}(\psi(2S)\rightarrow \gamma \eta_{c}(2S))$.  
}

\end{abstract}
\maketitle

\oddsidemargin  -0.2cm
\evensidemargin -0.2cm


\section{Introduction}

Charmonium states play an important role in understanding the strong interaction. Despite being known for more than half a century, the charmonium system still leaves much room for further study, especially in the spin singlet sector~\cite{ref_1}  which includes the $^{1}P_{1}$ state $h_c$ and the $2^{1}S_{0}$ state $\eta_c$(2S).
The $\eta_c(2S)$ is the first excited state of the pseudoscalar ground state $\eta_c$, lying just below the mass of its vector counterpart, $\psi(2S)$~\cite{ref_1}. It was first observed by the Belle collaboration~\cite{ref_2} in the process $B^{\pm} \rightarrow K^{\pm} \eta_c(2S)$ with $\eta_{c}(2S)\to K^{0}_{S}K^{\pm}\pi^{\mp}$. The $\eta_c(2S)$ was subsequently confirmed by BaBar~\cite{ref_3}, CLEO~\cite{ref_4}, and Belle~\cite{ref_5} in $\gamma\gamma$-fusion to $K\overline K \pi$ final state, by BaBar~\cite{ref_6} in double charmonium production process $e^+e^-\to J/\psi c\bar{c}$ and Belle~\cite{ref_7} in the inclusive process $e^+e^-\to J/\psi + anything$. BESIII studied $\psi(2S)\rightarrow\gamma\eta_c(2S)$ with $\eta_c(2S)\rightarrow K^{0}_{S}K^{\pm}\pi^{\mp}$ and $\eta_c(2S)\rightarrow K^{+}K^{-}\pi^{0}$~\cite{WangLL}, and showed consistent results with  previous experimental measurements and theoretical calculations.

Exploring hadronic transitions between charmonium states offers a valuable opportunity
to study the dynamics of heavy quarks.
Experimental~\cite{G.S.} and theoretical ~\cite{Voloshin,LS} investigations have provided a comprehensive understanding of those processes involving the emission of two-pion in the low energy chiral dynamics framework. Such decays are described as that heavy quarkonium transition generates a soft gluonic field which then produces the light mesons in QCD picture. The two-pion transition between $^{1}S_{0}$ states of the charmonium have not yet been observed.

The transition amplitude of the $\eta_c(2S)\rightarrow\pi^{+}\pi^{-}\eta_c$ decay ~\cite{ref_9} is expected to have the same linear dependence $q^2$ as the $\psi(2S)\rightarrow \pi^{+}\pi^{-}J/\psi $ decay~\cite{JZ}, where $q$ is the total 4-momentum of the pion pair. By estimating only the phase space (PHSP) in the single-channel approach, the branching fraction of $\eta_c(2S)\rightarrow\pi^{+}\pi^{-}\eta_c$ is estimated to be $\sim 5\%$~\cite{ref_9,pdg}. This decay may be further suppressed due to the contribution of the chromo-magnetic interaction in the decay amplitude~\cite{revise}.

A previous study indicated that the $\eta_c(2S)\rightarrow\pi^{+}\pi^{-}\eta_c$ decay is the most probable decay mode of $\eta_c(2S)$~\cite{ref_9}. The BaBar collaboration set the 90\% confidence level (C.L.) upper limit of $\mathcal{B}(\eta_c(2S)\rightarrow\pi^{+}\pi^{-}\eta_c)\textless 7.4\%$ based on the $e^{+}e^{-}$ collision data, corresponding to an integrated luminosity of (429.1$\pm$1.9) fb$^{-1}$ collected at the $\Upsilon(4S)$ resonance~\cite{ref_8}. In recent years, the BESIII collaboration collected a large data sample of $\psi(2S)$, which provides a great opportunity to  search for $\eta_c(2S)\rightarrow\pi^{+}\pi^{-}\eta_c$ and validate the theoretical calculations~\cite{Voloshin,LS}. 

 To date, only a few decay modes of $\eta_{c}(2S)$ have been observed, therefore, searching for new decay modes is important for an in-depth understanding of its properties. 
 Using 106 million $\psi(2S)$ events collected by the BESIII detector, an evidence for the decay  $\eta_{c}(2S)\rightarrow\pi^{+}\pi^{-}K^{0}_{S}K^{\pm}\pi^{\mp}$  is found with a statistical significance of 
 greater than 4$\sigma$~\cite{yliang}.
With $(27.12\pm0.14)\times10^8$ $\psi(2S)$ events taken by the BESIII detector, which is about 25 times larger than the previous data sample~\cite{yliang}, we perform a search for $\eta_c(2S)\to\pi^+\pi^-\eta_c$ and an improved  measurement of the branching fraction of $\eta_{c}(2S)\rightarrow\pi^{+}\pi^{-}K^{0}_{S}K^{\pm}\pi^{\mp}$.

\section{BESIII Detector and Monte Carlo Simulation}
The BESIII detector~\cite{ref:bes3} records symmetric $e^+e^-$ collisions 
provided by the BEPCII storage ring~\cite{ref:bepc}
in the center-of-mass energy range from 2.0 to 4.95~GeV,
with a peak luminosity of $1 \times 10^{33}\;\text{cm}^{-2}\text{s}^{-1}$ 
achieved at $\sqrt{s} = 3.77\;\text{GeV}$. 
BESIII has collected large data samples in this energy region~\cite{BESIII:2020nme} \cite{Zhang:2022bdc}.
The cylindrical core of 
the BESIII detector consists of a helium-based main drift chamber (MDC), a plastic 
scintillator time-of-flight (TOF) system, and a CsI(Tl) electromagnetic calorimeter (EMC), 
which are all enclosed in a super conducting solenoidal magnet providing a 1.0~T magnetic field. The solenoid is supported by an octagonal flux-return yoke with 
resistive plate counter modules interleaved with steel for muon identification. The acceptance for charged particles and photons is 93\% of the full solid angle, and the charged-particle 
momentum resolution at 1~GeV/c is 0.5\%. The photon energy resolution is 2.5\% (5\%) at 1.0 
GeV in the barrel (end-cap) region.
The time resolution in
the TOF barrel region is 68 ps, while that in the end-cap region was 110 ps. The end-cap TOF system was upgraded in 2015 using multigap resistive plate chamber technology, 
providing a time resolution of 60~ps,
which benefits $\sim$84\% of the data used in this analysis~\cite{Li:2017jpg,Guo:2017sjt,Cao:2020ibk}.

Simulated data samples produced with a \textsc{geant4}-based  Monte Carlo (MC) 
 package~\cite{GEANT4:2002zbu}, which includes the geometric 
 description of the BESIII detector and the detector response, are used to determine 
 detection efficiencies and to estimate backgrounds. The simulation models the beam-energy 
 spread and initial-state radiation (ISR)  in  $e^{+}e^{-}$ annihilations with the 
 generator \textsc{kkmc}~\cite{Jadach:1999vf,Jadach:2000ir}. 
An inclusive MC sample of  2.7 billion  $\psi(3686)$ events is used to investigate potential background. The inclusive MC sample includes the production 
 of the  $\psi(3686)$ resonance, the ISR production of the $J/\psi$, and
the continuum processes incorporated in {\sc
kkmc}. The known particle decays are modelled with 
 \textsc{evtgen}~\cite{Lange:2001uf,Ping:2008zz}  using branching fractions taken from the 
 Particle Data Group~\cite{pdg}, while the remaining unknown decays are estimated with \textsc{lundcharm}~\cite{Chen:2000tv,Yang:2014vra}.

To estimate the signal selection efficiency, detector resolution and background, several exclusive MC samples are generated.  Specific generators  are employed for channels $\psi(2S)\to\gamma\eta_{c}(2S)$ and $\gamma \chi_{c1,2}$, which are based on specialized models that have been packaged and customized specifically for BESIII detector. The polar angle ($\theta_{\gamma}$) of the radiation photon in the rest frame of $\psi(2S)$ follows (1 + $\lambda\cos^{2}\theta_{\gamma}$) function, where the value of $\lambda$ is 1 for $\eta_{c}(2S)$, -1/3 for $\chi_{c1}$ and 1/13 for $\chi_{c2}$, respectively~\cite{Tanenbaum:1977eg}. Two decay modes $\eta_{c}\to K^{+}K^{-}\pi^{0}$ and $\eta_{c}\to K^{0}_{S}K^{\pm}\pi^{\mp}$ are generated according to the Dalitz distribution measured by BESIII~\cite{etac}.
The exclusive MC samples  $\psi(2S)\to\gamma\pi^{+}\pi^{-}K^{0}_{S}K^{\pm}\pi^{\mp}$  $(\gamma\pi^{+}\pi^{-} K^{+}K^{-}\pi^{0})$,  $\psi(2S)\to\pi^{0}\pi^{+}\pi^{-}K^{0}_{S}K^{\pm}\pi^{\mp}$ $(\pi^0\pi^+\pi^-K^{+}K^{-}\pi^{0})$, $\psi(2S)\to(\gamma_{\mathrm{FSR}})\pi^{+}\pi^{-}K^{0}_{S}K^{\pm}\pi^{\mp}$ $((\gamma_{\mathrm{FSR}})\pi^{+}\pi^{-} K^{+}K^{-}\pi^{0})$ (here, $\gamma_{\mathrm FSR}$ stands for a photon from final state radiation), are generated uniformly in the phase space(PHSP).

The data sample collected at the center-of-mass energies of 3.65 GeV, corresponding to integrated luminosity of 410 pb$^{-1}$, is used to estimate the non-resonant continuum background contributions.

\section{Measurement of \boldmath{$\eta_{ \MakeLowercase{c}}(2S)\to\pi^{+}\pi^{-}\eta_{ \MakeLowercase{c}}$}}\label{ppetac1}

\subsection{Event selection}\label{evt_selection}

In this analysis, $\eta_c$ is reconstructed with $K_S^0K^{\pm}\pi^{\mp}$ and $K^{+}K^{-}\pi^{0}$, where  $K_S^0$ ($\pi^{0}$) is reconstructed with pairs of $\pi^+\pi^-$ ($\gamma \gamma$). 

Charged tracks are reconstructed in the MDC using good helix fits, and are required to satisfy $|\rm{cos}\theta|<0.93$, where $\theta$ is the polar angle with respect to the $z$-axis which is the symmetry axis of the MDC. For charged tracks apart from the $K^0_S$ decays, the distance of the closest approach to the interaction point must be less than 10 cm along the $z$-axis, and less than 1 cm in the transverse plane. 
By combining the $dE/dx$ and TOF information, the $\chi^2_{\rm PID}(i)$ ($i = K, \pi$, or $p$) is calculated for each charged track for each hadron $i$ hypothesis. Both PID and kinematic fit information are used to
determine the particle type of each charged track, as discussed below.

 

Photon candidates are identified using showers in the EMC.
 The deposited energy of each shower must be greater than 25~MeV in the barrel region ($|\rm{cos}\theta|<0.80$) and in the end-cap region ($0.86<|\rm{cos}\theta|<0.92$). To reduce electronic noise and showers unrelated to the event, the difference between the EMC time and the event start time is required to be within [0, 700] ns. The angle subtended by the EMC shower and the position of the closest extrapolated charged track at the EMC must be greater than 10 degrees. Candidate events must have at least one good photon (three good photons) for the decay mode $\eta_c\to K^0_S K^{\pm}\pi^{\mp}$ ($\eta_c\to K^+K^-\pi^0$).
 
$\pi^{0}$ Candidates are selected with the invariant mass of $\gamma \gamma$ pair satisfying $|M_{\gamma\gamma}-m_{\pi^0}|<0.015$~GeV/$c^{2}$, where $m_{\pi^{0}}$ is the known mass of $\pi^{0}$~\cite{pdg}.  To improve the energy resolution, a one-constraint (1C) kinematic fit is performed with the $\pi^0$ mass constraint.
A kinematic constraint between the production and decay vertices, and a second vertex fit algorithm based on the least square method are employed for $K^0_S$ candidate. The $K^0_{S}$ candidate must be satisfied with a decay length more than 2 standard deviations of the vertex resolution away from the interaction point, and its invariant mass to be within 0.02~GeV/$c^{2}$ of the $K^0_S$ known mass~\cite{pdg}.  
The events with $|M_{K^{0}_{S}K^{\pm}\pi^{\mp}(K^{+}K^{-}\pi^{0})}-m_{\eta_{c}}|<0.05$~GeV/$c^{2}$  are setelcted for further analysis,
where $M_{K^{0}_{S}K^{\pm}\pi^{\mp}(K^{+}K^{-}\pi^{0})}$ is the invariant mass of $K^{0}_{S}K^{\pm}\pi^{\mp} (K^{+}K^{-}\pi^{0})$ and $m_{\eta_{c}}$ is the known mass of $\eta_{c}$ meson~\cite{pdg}.
      
For the process of $\eta_{c}(2S)\rightarrow \pi^{+}\pi^{-}\eta_{c}$, $\eta_{c} \rightarrow K^{+}K^{-}\pi^{0}$, a five-constraints (5C) kinematic fit is performed with the constraints of total four-momentum conservation and two photons with the invariant mass constrained to the known $\pi^{0}$ mass.
For the process of $\eta_{c}(2S)\rightarrow \pi^{+}\pi^{-} \eta_{c}, \eta_{c} \rightarrow K^{0}_{S}K^{\pm}\pi^{\mp}$, a four-constraints~(4C) kinematic fit is performed.
The $\chi^2_{\rm 5C}<25$ and $\chi^2_{\rm 4C}<40$ are further required for the 
decay mode $\eta_{c} \rightarrow K^{+}K^{-}\pi^{0}$ and $\eta_{c} \rightarrow K^{0}_{S}K^{\pm}\pi^{\mp}$, respectively. These values are 
determined by optimizing the Figure-of-Merit (FOM) $S/\sqrt{S+B}$, where $S$ represents the number of signal events from MC simulation, and $B$ represents the number of normalized background events estimated from the left-hand sideband region of $\eta_c$ in the data.

%

     For each event, if there are multiple combinations satisfying the above criteria, only the combination with the minimum $\chi^2_{\rm tot} = \chi^2_{\rm 4C}+\chi^2_{\rm 1C}+\chi^2_{\rm PID}+\chi^2_{\rm vertex}$ is kept, where $\chi^2_{\rm{4C}}$ is from the 4C kinematic fit for four-momentum conservation, $\chi^2_{\rm 1C}$ is from the 1C kinematic fit for $\pi^0$ mass (this term is considered only in $\eta_{c} \rightarrow K^{+}K^{-}\pi^{0}$ mode), $\chi^2_{\rm PID}$ is the sum of the $\chi^2_{\rm PID}(i)$ for each charged track in the event, and $\chi^2_{\rm vertex}$ is from the $K_S^0$ secondary-vertex fit (this term is only taken into account in the $\eta_{c} \rightarrow K^{0}_{S}K^{\pm}\pi^{\mp}$ mode).      
       
      The background events from $\psi(2S)\to\pi^{+}\pi^{-}J/\psi$ process in both channels are suppressed by requiring the recoil mass of $\pi^{+}\pi^{-}$ to be less than 3.07~GeV$/c^{2}$. The background events from the $\psi(2S)\to \omega K^{+}K^{-}$ in the $\eta_{c}(2S)\rightarrow\pi^{+}\pi^{-}\eta_{c},\eta_{c}\rightarrow K^{+}K^{-}\pi^{0}$ decay is excluded by requiring $|M_{\pi^{0}\pi^{+}\pi^{-}}-m_\omega |\textgreater$0.02~GeV/$c^{2}$, where $M_{\pi^{0}\pi^{+}\pi^{-}}$ is the invariant mass of $\pi^{0}\pi^{+}\pi^{-}$ and $m_\omega$ is the known mass of $\omega$~\cite{pdg}.


\subsection{Background estimation}\label{bkg_etac}
A detailed study of the inclusive MC sample indicates that 
the backgrounds for $\eta_c(2S)\rightarrow\pi^{+}\pi^{-}\eta_c$, $\eta_c \rightarrow K^0_SK^{\pm}\pi^{\mp}$ process mainly come from
(I) $\psi(2S)\rightarrow (\gamma_{\mathrm{FSR}})\pi^{+}\pi^{-} K^0_S K^{\pm}\pi^{\mp}$, where $\gamma_{\rm FSR}$ stands for a photon from final state radiation (FSR), 
(II)  $\psi(2S)\rightarrow\gamma\eta_c(2S),  \eta_c(2S)\rightarrow\pi^{+}\pi^{-} K^0_S K^{\pm} \pi^{\mp}$, and (III) $\psi(2S)\rightarrow\gamma\chi_{c2}, \chi_{c2}\rightarrow\pi^{+}\pi^{-} K^0_S K^{\pm} \pi^{\mp}$.
The background events in the $\eta_c(2S)\rightarrow\pi^{+}\pi^{-} \eta_c$, $\eta_c \rightarrow K^{+}K^{-}\pi^{0}$ process mainly come from
(I) $\psi(2S)\rightarrow (\gamma_{\mathrm{FSR}})\pi^{+}\pi^{-} K^{+}K^{-} \pi^{0}$;
(II) $\psi(2S)\rightarrow \pi^{0}\pi^{+}\pi^{-}\eta_c, \eta_c \rightarrow K^{+}K^{-} \pi^{0}$;
(III)  $\psi(2S)\rightarrow\gamma\eta_c(2S),  \eta_c(2S)\rightarrow\pi^{+}\pi^{-}K^{+}K^{-}\pi^{0}$, and (IV) $\psi(2S)\rightarrow\gamma\chi_{c2}, \chi_{c2}\rightarrow\pi^{+}\pi^{-}K^{+}K^{-}\pi^{0}$.
 According to the studies of continuum data and inclusive MC sample, the contributions from continuum processes and events with misidentified $\pi$ or $K$ are found to be negligible.
The remaining background channels with the same final states as the signal make only a small contribution, and the corresponding lineshapes are smooth within the $\pi^{+}\pi^{-}K^{+}K^{-}\pi^{0}$ or $\pi^{+}\pi^{-}K^{0}_{S}K\pi$ invariant mass spectrum, which are described by the MC shape generated uniformly in PHSP.

\begin{figure}
    \centering
    \includegraphics[width=0.49\textwidth]{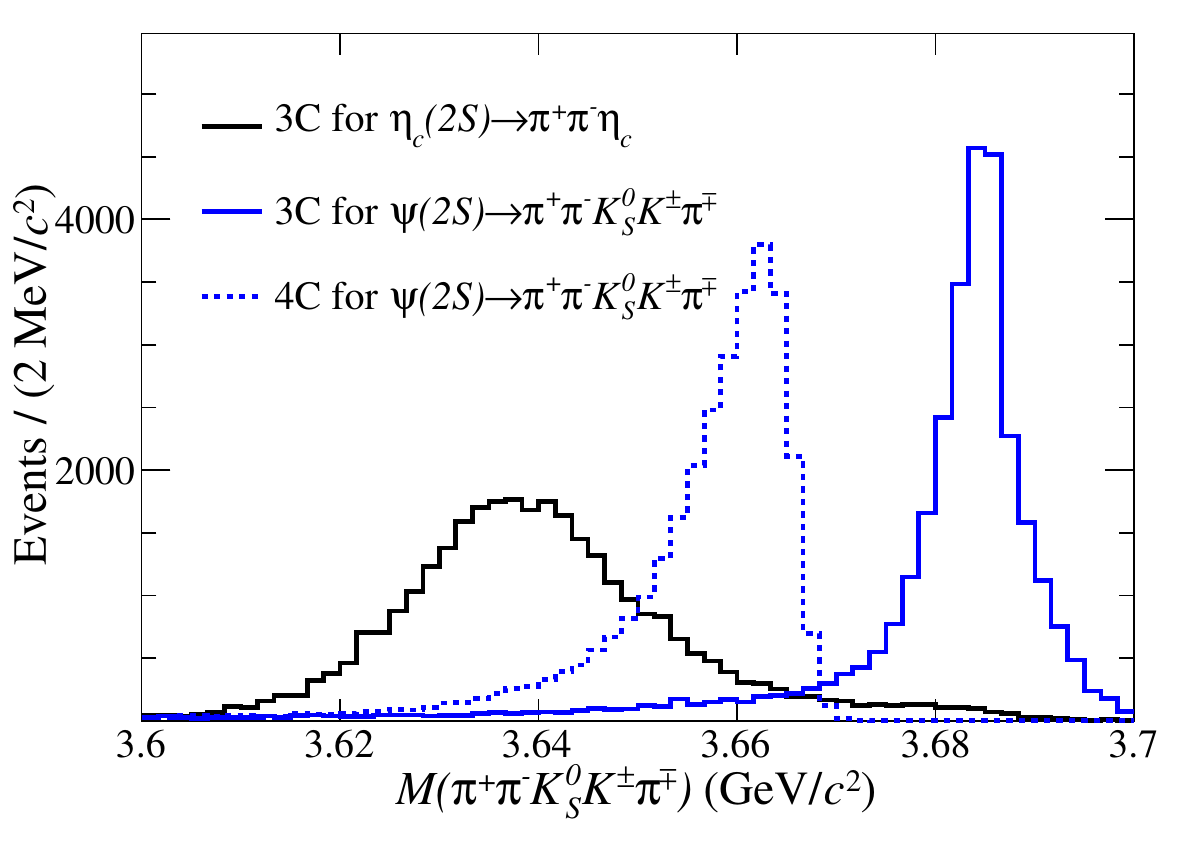}
    \caption{Distributions of the invariant mass spectra of $\pi^{+}\pi^{-} K^0_S K^{\pm}\pi^{\mp}$ for $\eta_{c}(2S)\rightarrow \pi^{+}\pi^{-} \eta_{c},  \eta_{c}\rightarrow K^{0}_{S}K^{\pm}\pi^{\mp}$ process with 3C (black solid line),  $\psi(2S)\rightarrow \pi^{+}\pi^{-} K^{0}_{S}K^{\pm}\pi^{\mp}$ process with 3C (blue solid line) and 4C kinematic fit (blue dashed line).}
    \label{kskpi_fp}
\end{figure}

The backgrounds from $\psi(2S)\rightarrow \pi^{+}\pi^{-} K^0_S K^{\pm} \pi^{\mp}$ and $\psi(2S)\rightarrow \pi^{+}\pi^{-} K^{+} K^{-} \pi^{0}$, either with a fake photon or a real soft photon from FSR, produce a peak near the $\eta_{c}(2S)$ signal on the invariant mass spectrum of $\pi^{+}\pi^{-}K^{0}_{S}K^{\pm}\pi^{\mp}$ ($\pi^{+}\pi^{-}K^{+}K^{-}\pi^{0}$).
This peak is shifted to a higher mass region after applying the three-constraints kinematic fit (3C) in which the energy of the M1 photon (M1 represents the magnetic dipole transition, the leading order transition amplitude.~\cite{Eichten:2007qx}) candidate is allowed to freely float~\cite{WangLL}. The 3C fit keeps a similar mass resolution as the 4C fit and has better signal-background separation, as shown in Fig.~\ref{kskpi_fp}, therefore, the result from the 3C fit is taken as the final mass spectrum. The contribution of such background is connected to the ratio $f_{\mathrm{FSR}}$, which is defined as
\begin{equation}
\label{FSR_eq}
f_{\mathrm{FSR}}= \frac{N_{\mathrm{FSR}}}{N_{\mathrm{nonFSR}}} \,  ,\\
\end{equation}
where $N_{\mathrm{FSR}}$ and $N_{\mathrm{nonFSR}}$ refer to the number of FSR events and non-FSR events, respectively. The relative ratio $f_{\mathrm{FSR}}$ between data and MC simulation is estimated to be $1.62\pm0.13$ using the control sample  $\psi(2S)\to\gamma\chi_{c0},\chi_{c0}\to3(\pi^{+}\pi^{-})(\gamma_{\mathrm{FSR}})$~\cite{hex_pi}.

\linenumbers

  The contributions of peaking backgrounds from $\psi(2S)\to\gamma\eta_{c}(2S), \eta_{c}(2S)\rightarrow \pi^{+}\pi^{-}K^{0}_{S}K^{\pm}\pi^{\mp}$ and $\eta_{c}(2S)\rightarrow \pi^{+}\pi^{-}K^{+}K^{-}\pi^{0}$ processes are determined by
    \begin{linenomath*}
    \begin{equation}
         N^{\rm peak}_{\rm bkg}= N_{\psi(2S)}\times \mathcal{B}_{1}\times \mathcal{B}_{2}\times \varepsilon ,
    \end{equation}
    \end{linenomath*}
where $N_{\psi(2S)}$ represents the total number of $\psi(2S)$ events, $\varepsilon$ = 4.9$\times10^{-3}$ (2.0$\times10^{-3}$) is the detection efficiency for $\eta_{c}(2S)\to \pi^{+}\pi^{-}K^{0}_{S}K^{\pm}\pi^{\mp} $ $(\pi^{+}\pi^{-}K^{+}K^{-}\pi^{0})$ mode. 
$\mathcal{B}_{1}$ is the branching fraction of $\psi(2S)\to \gamma  \eta_{c}(2S)$ taken from the PDG~\cite{pdg}. $\mathcal{B}_{2}$ is the branching fraction of  $\eta_{c}(2S)\to \pi^{+}\pi^{-}K^{0}_{S}K^{\pm}\pi^{\mp} $ $(\pi^{+}\pi^{-}K^{+}K^{-}\pi^{0})$, which is estimated to be $(1.33\pm1.04)\times10^{-2}$, as discussed in Sec.~\ref{noetac} ($(1.4\pm1.0)\times{10^{-2}}$ cited from the PDG~\cite{pdg}).
With these inputs, the numbers of peaking background events $N^{\rm peak}_{\rm bkg}$ are determined to be 122$\pm$36 and 54$\pm 39$ for $\eta_{c}(2S)\to\pi^{+}\pi^{-}K^{0}_{S}K^{\pm}\pi^{\mp}$ and  $\eta_{c}(2S)\to\pi^{+}\pi^{-}K^{+}K^{-}\pi^{0}$ mode, respectively.

The background of $\psi(2S)\rightarrow \pi^{+}\pi^{-}K^{+}K^{-}\pi^{0}\pi^{0}$ contributes a smooth component in the invariant mass spectrum of $\pi^{+}\pi^{-}K^{+}K^{-}\pi^{0}$ that is described by the MC-simulated shape.

\subsection{Signal yield}\label{sig_extr}

We simultaneously fit the $\pi^{+}\pi^{-}K^{0}_{S}K^{\pm}\pi^{\mp}$ and $\pi^{+}\pi^{-}K^{+}K^{-}\pi^{0}$ invariant mass distributions with an unbinned maximum likelihood method to determine the number of signal events of $\eta_{c}(2S)\rightarrow\pi^{+}\pi^{-}\eta_{c}$. 

The lineshape of $\eta_c$(2S) produced by the M1 transition can be effectively described as
\begin{linenomath*}
\begin{equation}
\left ( E_{\gamma}^{3}\times BW\left ( m \right )\times f_{d}(E_{\gamma}) \right) \bigotimes F_{\rm res} \times \varepsilon\left(m\right) \,,
\label{eq7}
\end{equation}
\end{linenomath*}
where $m$ is the invariant mass of $\pi^{+}\pi^{-}K^{0}_{S}K^{\pm}\pi^{\mp}$ $(\pi^{+}\pi^{-}K^{+}K^{-}\pi^{0})$, $E_{\gamma}= \frac{m_{\psi\left(2S\right)}^{2}-m^{2}}{2m_{\psi\left(2S\right)}}$ is the energy of the transition photon in the rest frame of $\psi(2S)$, $BW(m)$ is the Breit-Wigner function for $\eta_c$(2S). The mass and width of $\eta_{c}(2S)$ are fixed to 3643.4~MeV/$c^{2}$ and 19.80~MeV, respectively, which are taken from the previous BESIII measurement~\cite{hex_pi}. $f_{d}(E_{\gamma})$ is the function to damp the divergent tail raised by $E^3_\gamma$~\cite{KERD},
\begin{linenomath*}
 \begin{equation}
  \label{df}
f_{d}(E_{\gamma})= \frac{E^{2}_{0}}{E_{\gamma}E_{0}+\left ( E_{\gamma}-E_{0} \right )^{2}} \,,
 \end{equation}
 \end{linenomath*}
 where $E_{0}= \frac{m_{\psi(2S)}^{2}-m_{\eta_c(2S)}^{2}}{2m_{\psi(2S)}}$. $\varepsilon(m)$ is the mass dependent efficiency function which is estimated by MC simulation.  $F_{\rm res}$ is a double Gaussian function describing the detector resolution. 

 Events from $\psi(2S)\to\gamma\chi_{c2}, \chi_{c2}\to \pi^{+}\pi^{-}K^{0}_{S}K^{\pm}\pi^{\mp}/\pi^{+}\pi^{-}K^{+}K^{-}\pi^{0}$ are included in the fit. The lineshape of $\chi_{c2}$ is described by an MC shape convolved with a Gaussian function. The Gaussian function accounts for the difference in mass resolution between data and MC simulation, and its parameters are floated in the fit. The shape of the FSR processes, $\psi(2S)\to\gamma_{\mathrm{FSR}}\pi^{+}\pi^{-}K^{+}K^{-}\pi^{0}$ and $\psi(2S)\to\gamma_{\mathrm{FSR}}\pi^{+}\pi^{-}K^{0}_{S}K^{\pm}\pi^{\mp}$, as well as the  $\psi(2S)\to\gamma\chi_{c2}$ and $\psi(2S)\to \pi^{0}\pi^{+}\pi^{-}K^{+}K^{-}\pi^{0}$ processes are fixed, while these background yields are floated in the fit. The yield of peaking background $\eta_c(2S)\to \pi^+\pi^-K^0_S K^{\pm}\pi^{\mp}$ ($\pi^+\pi^-K^+K^-\pi^0$) is fixed in the fit according to estimation in Sec.~\ref{ppetac1} B. The remaining backgrounds are described by a third-order Chebychev polynomial function with both the parameters and number of events are floated. 
 
\begin{figure*}[htbp]
    \begin{center}
    \subfigure{
   \includegraphics[width=17.5cm,angle=0]{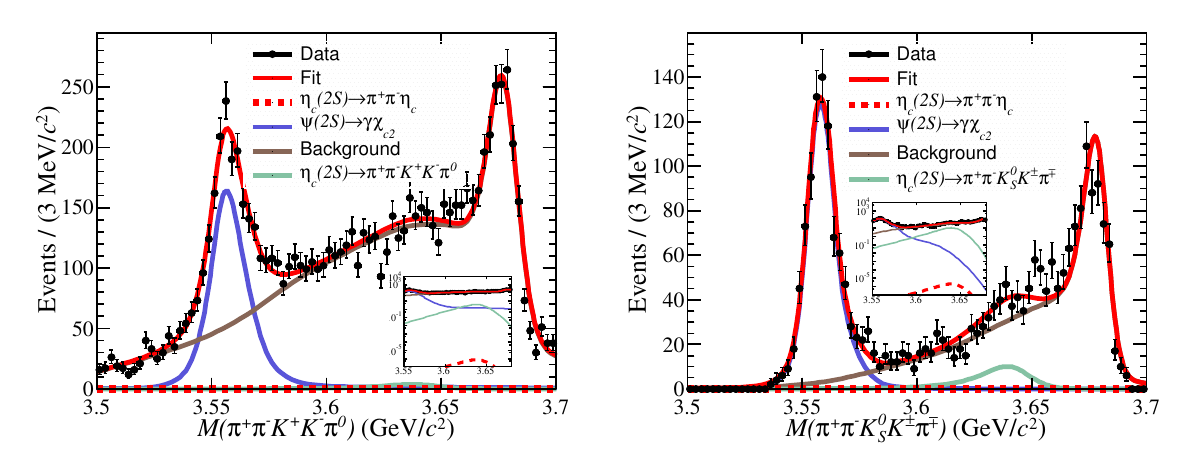} \put(-360,-10){(a)}\put(-110,-10){(b)}
   }   
    \end{center}
    \caption{ The results of the simultaneous  fit to the invariant mass distributions of (a) $M^{\rm 3C}_{\pi^{+}\pi^{-}K^{+}K^{-}\pi^{0}}$ and (b) $M^{\rm 3C}_{\pi^{+}\pi^{-}K^{0}_{S}K^{\pm}\pi^{\mp}}$, and the magnified view of the signal region on a logarithmic scale. The black dots with error bars are data, the blue solid lines are for $\psi(2S)\to \gamma\chi_{c2}$, the red dashed lines are the MC signals, the green solid lines are the peaking background components, the red solid lines are the total fit curves, and the brown solid lines represent the remaining backgrounds.  }\label{fit_result}
\end{figure*}

 No significant $\eta_{c}(2S)$ signal is observed in the $\pi^{+}\pi^{-}K^{+}K^{-}\pi^{0}$($\pi^{+}\pi^{-}K^{0}_{S}K^{\pm}\pi^{\mp}$) invariant mass spectra after 3C kinematic fit. The fit results are displayed in Fig.~\ref{fit_result}. The fit qualities are $\chi^{2}/ndf _{\pi^{+}\pi^{-}K^{+}K^{-}\pi^{0}} =1.59 $ and $\chi^{2}/ndf _{\pi^{+}\pi^{-}K^{0}_{S}K\pi} = 1.26 $, where $ndf$ stands for the number of degrees of freedom. 

\subsection{Upper limit on branching fraction  }

The product branching fraction of $\mathcal{B}(\psi(2S)\rightarrow \gamma \eta_c(2S))\times$ $\mathcal{B}(\eta_c(2S)\rightarrow\pi^{+}\pi^{-}\eta_c)$ is  calculated as
\begin{linenomath*}
    \begin{equation}\label{uplimit}
\mathcal{B}= \frac{N_{\rm sig}}{N_{\psi(2S)}\times (\varepsilon_{3}\times \mathcal{B}_{3}+\varepsilon_{4}\times \mathcal{B}_{4})} \,.
\end{equation}
\end{linenomath*}

Here, $ N _{\rm sig}$ is the upper limit on the signal yield, $\mathcal{B}_{3}$ ($\mathcal{B}_4$) represents the branching fraction of $\eta_{c}$ hadronic decay $\eta_{c}\rightarrow K^{+}K^{-}\pi^{0}$ ($\eta_c\to K^{0}_{S}K^{\pm}\pi^{\mp}$), and $\varepsilon_{3}$ ($\varepsilon_{4}$) is the  efficiency obtained with MC simulation, which is 11.4\% (12.0\%) for $K^{+}K^{-}\pi^{0}$ ($K^{0}_{S}K^{\pm}\pi^{\mp}$) mode. An upper limit at the 90\% C.L. for $\mathcal{B}(\psi(2S)\rightarrow \gamma \eta_c(2S))\times$ $\mathcal{B}(\eta_c(2S)\rightarrow\pi^{+}\pi^{-}\eta_c)$ is set using the Bayesian method~\cite{Baysian}. 
The systematic uncertainties detailed in Section~\ref{sys} are taken into account in the upper limit calculation. First, we select the likelihood that produces the largest upper limit with considering the additive systematic terms. To incorporate the multiplicative terms into branching fraction limit, the probability of the measured branching fraction $P$($\mathcal{B}$) being determined using a maximum likelihood fit, is convolved with a probability density function of sensitivity, which is defined as the denominator of Eq.~\ref{uplimit}. The probability density function of sensitivity is assumed to be a Gaussian function with central value $\hat{S}$ and standard deviation $\sigma_{s}$~\cite{smear}, which is given by
\begin{equation}
{
P^{\prime}(\mathcal{B})=\int_{0}^{\infty}P\left ( \frac{S}{\hat{S}}\mathcal{B}\right ) {\rm exp} \left [ \frac{-(S-\hat{S})^{2}}{2\sigma^{2}_{s}}  \right ]dS,
}
\label{eq}
\end{equation}
where $S$ and $\hat{S}$ refer to the sensitivity, $S$ is the variable of integration and $\sigma_{s}$ is its multiplicative systematics which is discussed in Section~\ref{sys}. 
By scanning $P^{\prime}$($\mathcal{B}$), the branching fraction upper limit at the 90\% C.L. is calculated to be 2.21$\times 10^{-5}$ via the following equation
\begin{linenomath*}
	\begin{equation}
	0.9= \int_{0}^{\mathcal{B}^{up}} P^{\prime}(\mathcal{B} )d\mathcal{B} .
\end{equation}
\end{linenomath*}

\begin{figure}[htb]
\begin{center}
\includegraphics[width=0.47\textwidth]{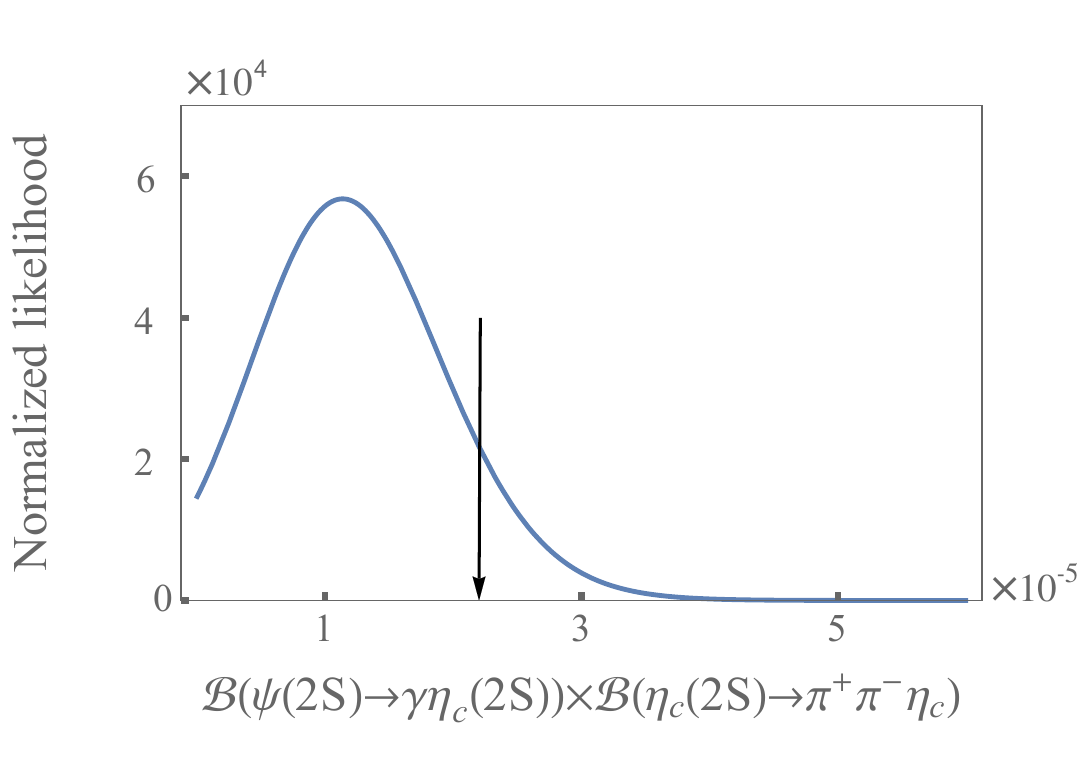}
\caption{Normalized likelihood versus the $\mathcal{B}(\psi(2S)\to\gamma\eta_{c}(2S))\times\mathcal{B}(\eta_{c}(2S)\to\pi^{+}\pi^{-}\eta_{c})$ after taking into account the systematic uncertainties. The black arrow corresponds to the product branching fraction at the 90\% C.L. upper limit.}
\label{upperL}
\end{center}
\end{figure}

The normalized likelihood distribution as a function of $\mathcal{B}(\psi(2S)\to\gamma\eta_{c}(2S))\times\mathcal{B}(\eta_{c}(2S)\to\pi^{+}\pi^{-}\eta_{c})$, taking into account the systematic uncertainties is shown in Fig.~\ref{upperL}.

    \section{MEASUREMENT OF \boldmath{$\eta_{ \MakeLowercase{c}}(2S)\to\pi^{+}\pi^{-}K^{0}_{S}K^{\pm}\pi^{\mp}$}}\label{noetac}
    
    \subsection{Event selection}
    
The selection criteria for $\eta_{c}(2S)\to\pi^{+}\pi^{-}K^{0}_{S}K^{\pm}\pi^{\mp}$ decay are almost the same as those for $\psi(2S)\to\gamma\eta_{c}(2S), \eta_{c}(2S)\to\pi^{+}\pi^{-}\eta_{c}, \eta_{c}\to K^{0}_{S}K^{\pm}\pi^{\mp}$, as described in Section~\ref{evt_selection}. But, the selection criteria for $\eta_{c}$ is removed. A 4C kinematic fit is performed under the $e^+e^-\rightarrow\gamma \pi^{+}\pi^{-}\pi^{+}\pi^{-}K^{\pm}\pi^{\mp}$hypothesis, and $\chi^{2}_{\rm 4C}$ is further required to be less than 25.
      
      \subsection{Background estimation}
      
The analysis of the inclusive MC sample shows that the backgrounds mainly come from (I) $\psi(2S)\rightarrow (\gamma_{\mathrm{FSR}})\pi^{+} \pi^{-} K^0_S K^{\pm}\pi^{\mp}$, (II) $\psi(2S)\rightarrow \pi^{0} \pi^{+}\pi^{-} K^0_S K^{\pm} \pi^{\mp}$, (III) $\psi(2S)\to\gamma \chi_{c1,2},\chi_{c1,2}\to\pi^+\pi^- K^0_S K^{\pm}\pi^{\mp}$, (IV) continuum process, and (V) events with pion or kaon misidentified. 
      
\begin{figure}
          \centering
          \includegraphics[width=0.49\textwidth]{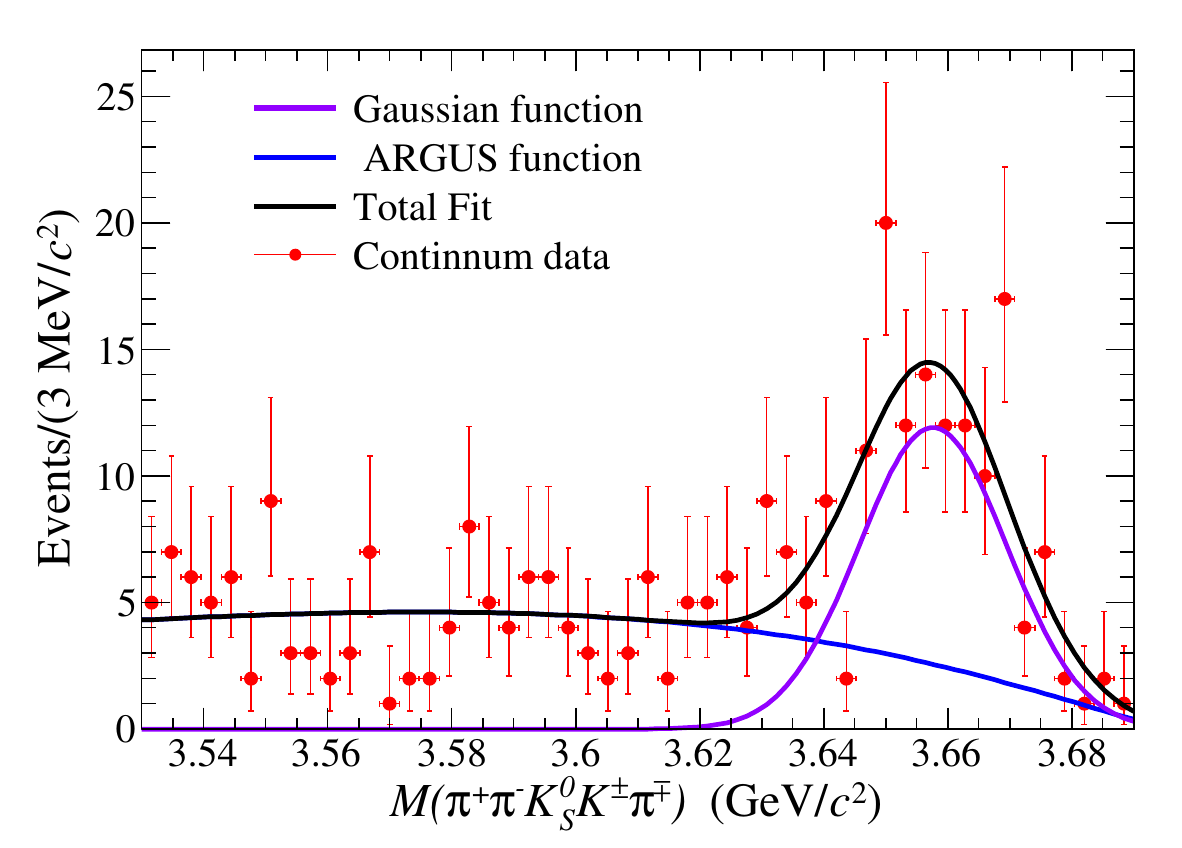}
          \caption{Distribution of $M^{\rm 3C}_{\pi^{+}\pi^{-}K^{0}_{S}K^{\pm}\pi^{\mp}}$. The purple line is the Gaussian function, the blue line is the ARGUS function, the red dots with error bars are continuum data and the black line is the total fit curve.}
          \label{cont}
      \end{figure}

The lineshaps for the invariant mass of $\pi^+\pi^-K^0_S K^{\pm}\pi^{\mp}$ are descirbed with the corresponding MC shapes for background process of (I) and (II). 
The lineshape of $\chi_{c1,2}$ is described with the MC shape convoluted with a Gaussian function for the difference in resolution of $M(\pi^+\pi^-K^0_S K^{\pm}\pi^{\mp})$ between data and MC simulation for the process of (III), while the number of events of each component and parameters in the Gaussian function are floated in the final fit. 
The data sample collected at $\sqrt{s}=3.65$ GeV is used to estimate the background from the continuum process $e^+e^-\to \gamma \pi^+\pi^- K^0_S K^{\pm}\pi^{\mp}$.
      The momenta and energies of final state particles are scaled to account for the difference in center-of-mass energy. The events passed the signal selection criteria are then normalized according to differences in integrated luminosity and cross section. The normalization factor is determined to be 9.8.      
To reduce the influence of statistical fluctuations, we model the shape of the background from the continuum process by the ARGUS function together with a Gaussian function, as shown in Fig.\ref{cont}, and the  number of events considering the normalization factor is fixed in the final fit.  We describe all other backgrounds with the shapes fixed to those from the inclusive MC sample, while the number of events is floated in the final fit.

      \subsection{Signal yield}
     The 4C kinematic fit is utilized in event selection, while the $M^{\rm 3C}_{\pi^{+}\pi^{-}K^{0}_{S}K\pi}$ distribution of the 3C kinematic fit is used for the signal measurement due to its improved signal-background separation. An unbinned maximum likelihood fit is performed. The lineshape of $\eta_c$(2S) is described by Eq.~\ref{eq7}.       

\begin{figure*}[htbp]
  \centering
  \subfigure{
  \label{noetac_a}
  	\includegraphics[width=0.49\textwidth]{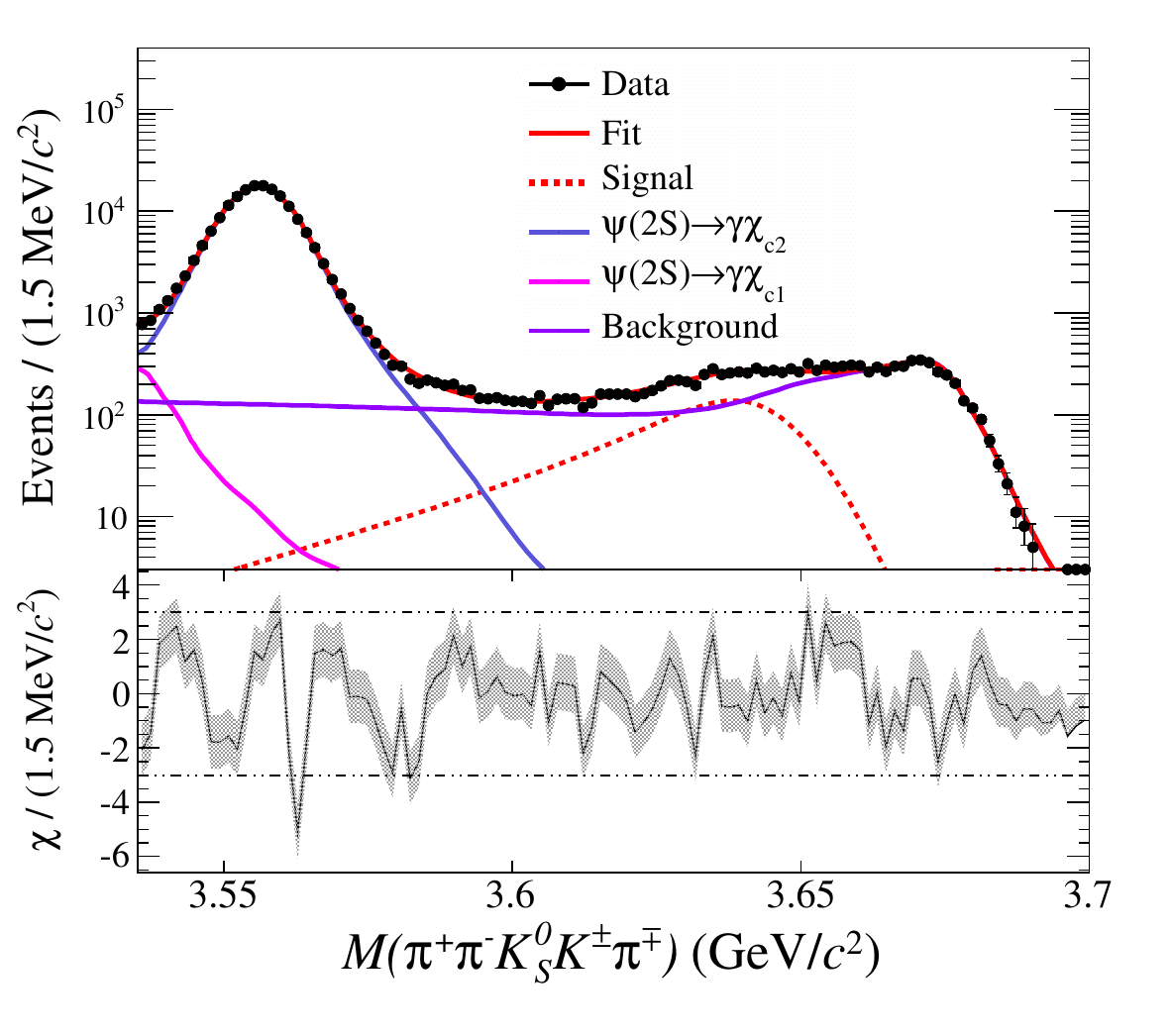}\put(-60,190){(a)}}%
  \subfigure{
  \label{noetac_b}
    \includegraphics[width=0.49\textwidth]{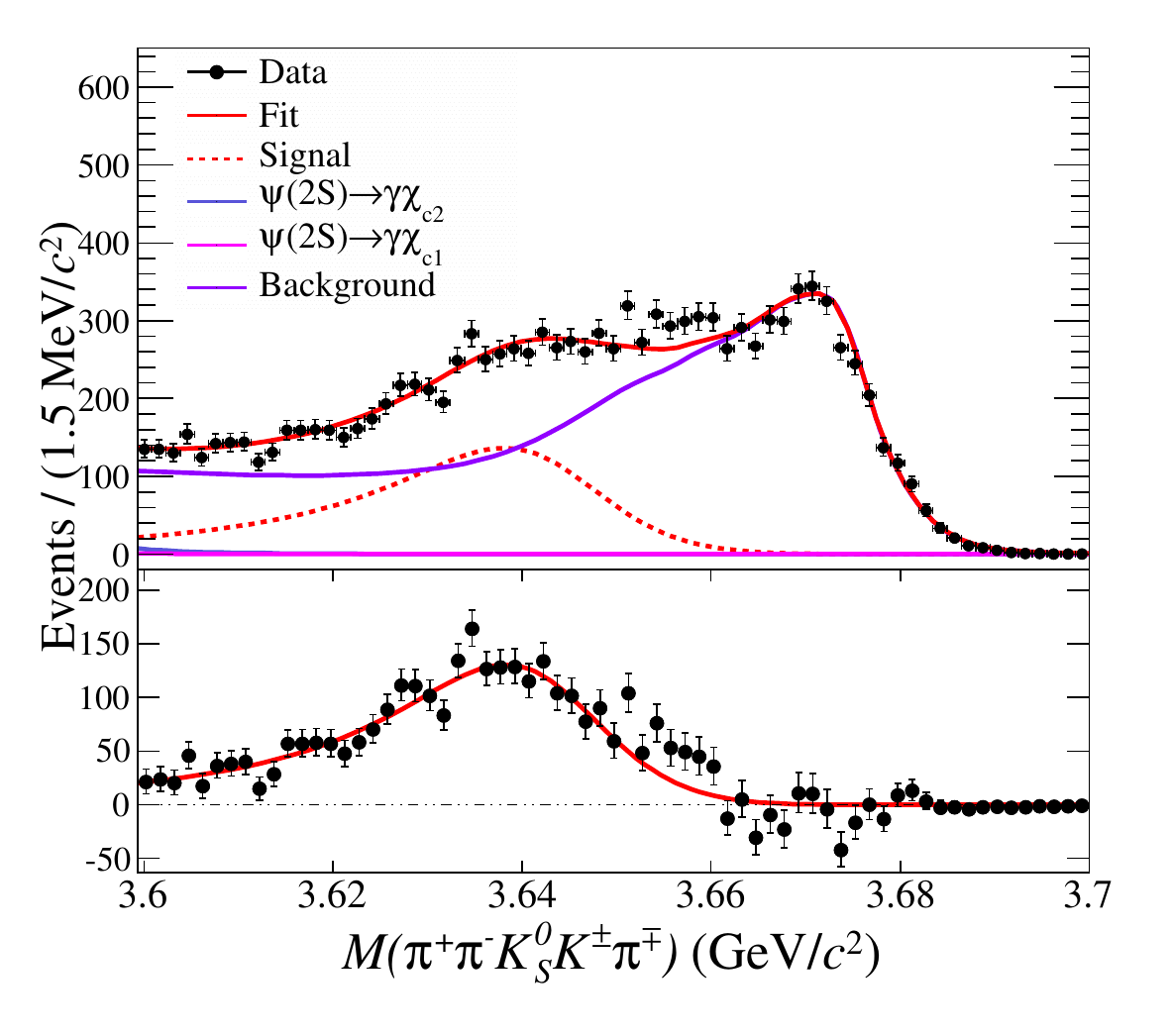
}\put(-60,190){(b)}} %

   \caption{(a) Distribution of $M^{\rm 3C}_{\pi^{+}\pi^{-}K^{0}_{S}K\pi}$, the related fit result and the pull distribution. (b) Zoomed  $M^{\rm 3C}_{\pi^{+}\pi^{-}K^{0}_{S}K\pi}$ spectrum and background residuals. The black dots with error bars are data, the blue solid line is the lineshape of $\psi(2S)\to\gamma\chi_{c2}$, the pink solid line is the lineshape of $\psi(2S)\to\gamma\chi_{c1}$, the purple solid line is for the remaining background, the red dotted line is the signal, and the red solid line is the total fit result.}
\end{figure*}

   Figure~\ref{noetac_a} displays all the different components involved in the fit process, along with the relative pull distribution. The zoomed distribution of the fit result and the background-subtracted distribution in the $\eta_{c}(2S)$ signal region are depicted in Fig.~\ref{noetac_b}. The yield of $\eta_{c}(2S)$, $N^{\prime}_{sig}$, is determined to be 3140 $\pm$ 241, with a  statistical significance of 10$\sigma$ calculated from the difference of the logarithmic likelihoods ($-$2($\mathcal{L}_{\rm sig}-\mathcal{L}_{\rm bkg}$)) taking into account the difference in the $ndf$ ~\cite{like_ref}. Here, $\mathcal{L}_{\rm sig}$ and $\mathcal{L}_{\rm bkg}$ are the negative logarithmic maximized likelihoods with and without a signal component, respectively.

\subsection{Branching fraction}

The product branching fraction $\mathcal{B}(\psi(2S)\to\gamma\eta_{c}(2S))\times\mathcal{B}(\eta_{c}(2S)\to\pi^{+}\pi^{-}K^{0}_{S}K^{\pm}\pi^{\mp})$ is calculated as
\begin{equation}
\mathcal{B}_{\rm pro}=\frac{N^{\prime}_{\rm sig}}{N_{\psi(2S)} \times  \varepsilon^{\prime} }  ,
\end{equation}
where $\varepsilon^{\prime}=12.4\%$ is the detection efficiency estimated by MC simulation.
The branching fraction of $\eta_{c}(2S)\to\pi^{+}\pi^{-}K^{0}_{S}K^{\pm}\pi^{\mp}$ is determined  to be (1.33$\pm$0.11) $\times 10^{-2}$, where the uncertainty is statistical only.

    \section{SYSTEMATIC UNCERTAINTY}\label{sys}
    
\subsection{Systematic uncertainty for \boldmath{$\eta_{c}(2S)\rightarrow \pi^{+}\pi^{-}\eta_{c}$}}\label{sysA}

The sources of systematic uncertainties are divided into two categories: additive and multiplicative terms.

 The additive uncertainties arise from the fit range, signal shape and background estimation. 
 A conservative approach is adopted by considering all possible variations and selecting the one that yields the largest upper limit. 

\begin{itemize}

\item[(i)] The systematic uncertainty arising from the fit range is determined by changing the invariant mass fit range from $[3.5,3.7]$~GeV/$c^{2}$ to $[3.52,3.72]$~GeV/$c^{2}$ and $[3.51,3.71]$~GeV/$c^{2}$.

  \item[(ii)]The uncertainty associated with the $\eta_c$ shape is estimated by changing damping function from Eq.~\ref{df} to an alternative one used by the CLEO Collaboration~\cite{CLEO}, while the mass parameter of the $\eta_{c}$(2S) is varied by $\pm 1 \sigma$. The systematic uncertainty caused by the resolution difference between data and MC simulation in the decay $\eta_{c}(2S)\rightarrow \pi^{+}\pi^{-}\eta_{c}$ is estimated by using the control samples $\chi_{c2}\rightarrow\pi^{+}\pi^{-}K^{+}K^{-}\pi^{0}$ and $\chi_{c2}\rightarrow\pi^{+}\pi^{-}K^{0}_{S}K^{\pm}\pi^{\mp}$.   
  The obtained mass resolution instead of the nominal one is used in the Gaussian function to describe the $\eta_c(2S)$ shape. 


 \item[(iii)]The uncertainty due to the background shape is estimated by altering the function from the third-order to the second-order Chebyshev polynomial function.  The relative ratio $f_{\rm FSR}$ is changed by $\pm 1\sigma$ to evaluate the systematic uncertainty due to FSR. The uncertainty due to peaking background is estimated by varying the number of $\eta_{c}(2S)\rightarrow \pi^{+}\pi^{-}K^{0}_{S}K^{\pm}\pi^{\mp}$ and $\eta_{c}(2S)\rightarrow \pi^{+}\pi^{-}K^{+}K^{-}\pi^{0}$  events by $\pm$1$\sigma$ for $\eta_c\to K^0_S K^{\pm}\pi^{\mp}$ and $\eta_c\to K^+K^-\pi^0$ decay mode, respectively.
 

\end{itemize}
 
The multiplicative uncertainties are involved in the determination of efficiency and branching fraction. These uncertainties include the following sources.
\begin{itemize}

\item[(i)]The uncertainty of the total number of $\psi(2S)$ events is estimated to be 0.5\% ~\cite{number}.

\item[(ii)]The uncertainty due to tracking efficiency for charged track is
determined to be 1.0\% per track using the control samples  $J/\psi\rightarrow\pi^{0}\pi^{+}\pi^{-}, J/\psi\rightarrow p \bar{p}\pi^{+}\pi^{-}$ and $J/\psi\rightarrow K^{0}_{S}K^{\pm}\pi^{\mp}+ c.c. $ ~\cite{pi_sys}.

\item[(iii)]
The systematic uncertainty due to photon detection efficiency is 1.0\% per photon, which is determined from the control samples $J/\psi \rightarrow \rho^{0}\pi^{0}$ and $e^{+}e^{-}\rightarrow\gamma\gamma$~\cite{gm}.

\item[(iv)]

The systematic uncertainty due to $K^0_S$ reconstruction is studied using control samples $J/\psi\to K^{*\pm}K^{\mp}$ and $J/\psi\to \phi K^{0}_{S}K^{\pm}\pi^{\mp}$, and estimated to be 1.0\%.

\item[(v)]Using a high purity control sample of $J/\psi\to\pi^{0}\Bar{p}p$, the systematic uncertainty from the $\pi^{0}$ reconstruction is determined to be 1.0\%~\cite{pi0_rec}.

\item[(vi)]The systematic uncertainty associated with the kinematic fit is assigned as the difference between the detection efficiencies before and after the helix parameter corrections~\cite{helix_par} in the MC simulation, which is 2.0\% for $\eta_{c}(2S)\to\pi^{+}\pi^{-}\eta_{c}, \eta_{c}\to K^{0}_{S}K^{\pm}\pi^{\mp}$, and 2.5\% for $\eta_{c}(2S)\to\pi^{+}\pi^{-}\eta_{c},\eta_{c}\to K^{+}K^{-}\pi^{0}$ mode.


\item[(vii)] To evaluate the uncertainty due to the $J/\psi$ veto, we vary the requirement on the recoil mass of $\pi^+\pi^-$ pair by $\pm$10 MeV/$c^2$. The maximum deviation in the measured result is taken as the systematic uncertainty.

\item[(viii)]
The systematic uncertainty arising from the $\eta_c$ mass window is estimated by adjusting the mass window by $\pm$ 10 MeV/$c^2$. The maximum difference in the branching fraction with respect to the nominal value is taken as the systematic uncertainty.

\item[(ix)]To evaluate the uncertainty due to the $\omega$ veto, we change the mass window of $\omega$ by $\pm$ 1 MeV/$c^2$.
The maximum deviation in the measured result is taken as the systematic uncertainty.

\item[(x)]
The uncertainties from the branching fractions of $\psi(2S)\to\gamma\eta_c(2S)$,  $\eta_c \rightarrow K^{+}K^{-}\pi^{0}$ and $\eta_c \rightarrow K^{0}_{S}K^{\pm}\pi^{\mp}$ are 71.4\% ~\cite{pdg}, 13.6\% and 11.2\%~\cite{etac}, respectively.

\end{itemize}

  \begin{table*}[tb]
\centering
\caption{Multiplicative systematic uncertainties (in \%) in the measured branching fraction for $\eta_{c}(2S)\to \pi^{+}\pi^{-}\eta_{c}$.}
\label{multi_sys}
\begin{tabular}{l|c|c|c} 
\hline
\multicolumn{1}{c|}{Source}                   & \multicolumn{1}{c|}{$\eta_c$(2S)$\rightarrow \pi^{+}\pi^{-}\eta_c$,$\eta_c\rightarrow K^{0}_{S}K^{\pm}\pi^{\mp}$} &$\eta_c$(2S)$\rightarrow \pi^{+}\pi^{-}\eta_c$,$\eta_c\rightarrow K^{+}K^{-}\pi^{0}$  
& \multicolumn{1}{c}{$\sigma^{i}_{\rm sum}$}  \\ 
\hline\hline
$N_{\psi(2S)}$                                          & 0.5                                                                                     & 0.5                                                                                         & 0.5                                   \\ 
\hline
Tracking efficiency                         & 6.0                                                                                       & 4.0                                                                                            & 5.4                                       \\ 
\hline
Photon reconstruction                                  & 1.0                                                                                      & 3.0                                                                                             & 1.6                                       \\ 
\hline
$K^0_S$ reconstruction                                  & 1.0                                                                                        &      -                                                                                           & 0.7                                         \\ 
\hline
$\pi^0$ reconstruction                                  &    -                                                                                        & 1.0                                                                                             & 0.3                                         \\ 
\hline
Kinematic fit                                          & 2.0                                                                                        & 2.5                                                                                           & 2.1                                       \\ 
\hline
$J/\psi$ veto                                          & 4.4                                                                                      & 4.0                                                                                             & 3.3                                         \\ 
\hline
$\eta_c$ mass window                                  & 4.0                                                                                        & 4.0                                                                                             & 3.1                                       \\ 
\hline
$\omega$ veto                                          &      -                                                                                      & 0.1                                                                                           & 0.03                                       \\ 
\hline
$\eta_c$ decay                                         & 11.2                                                                                     & 13.6                                                                                          & 8.9                                       \\ 
\hline
Total                                         & \multicolumn{1}{c|}{14.2}                                                                      & \multicolumn{1}{c|}{15.8}                                                                           & 11.7                                      \\ 
\hline
$\mathcal{B} (\psi(2S) \rightarrow \gamma \eta_c(2S))$ & \multicolumn{3}{c}{71.4}                                                                                                                                                                                                              \\
\hline
\end{tabular}
\end{table*}

 Table~\ref{multi_sys} summarizes the multiplicative systematic uncertainties.  The  combined relative systematic uncertainty is calculated by~\cite{comb_sys}
 \begin{linenomath*}
\begin{equation}
    \sigma^{i}_{\rm sum}=\frac{\sqrt{(\omega_{1}\delta^{i}_{1})^{2} +(\omega_{2}\delta^{i}_{2})^{2}+2\omega_{1}\omega_{2}\rho^{i}_{12}\delta^{i}_{1}\delta^{i}_{2}}}{\omega_{1}+\omega_{2}} \,, 
    \label{multi_tot}
\end{equation}
\end{linenomath*}
where $\omega_{1}$ and $\omega_{2}$ are $\mathcal{B}(\eta_{c}\rightarrow K^{+}K^{-}\pi^{0})\times\varepsilon_{3}$ and $\mathcal{B}(\eta_{c}\rightarrow K^{0}_{S}K^{\pm}\pi^{\mp})\times\varepsilon_{4}$,
$\delta^{i}_{1}$ and $\delta^{i}_{2}$ are the corresponding multiplicative uncertainties for the two decay modes. The correlation coefficiency $\rho_{12}$ is taken as 0 for those items that are uncorrelated between the two decay modes, including $\eta_{c}$ mass window, $K^{0}_{S}$ reconstruction, $J/\psi$ veto, $\omega$ veto, and branching faction of $\eta_{c}$ decay. For other systematic effects that are correlated between the two modes,  $\rho_{12}$ is taken as 1. 
Finally, the total combined systematic uncertainty $\sigma_{\rm sum}$ is assigned as 
\begin{linenomath*}
\begin{equation}
    \sigma_{\rm sum}=\sqrt{\Sigma (\sigma^{i}_{\rm sum})^{2}} \,.
    \label{tot_sys}
\end{equation}
\end{linenomath*}
, which is listed in the last column in Table~\ref{multi_sys}.

The total multiplicative systematic uncertainty is the sum in quadrature of the individual components, and is considered by smearing the likelihood curve~\cite{smear}.

\subsection{Systematic uncertainties for \boldmath{$\eta_{c}(2S)\rightarrow \pi^{+}\pi^{-}K^{0}_{S}K\pi$}}

The systematic uncertainties in the measurement of the branching fraction of  $\eta_{c}(2S)\rightarrow \pi^{+}\pi^{-}K^{0}_{S}K\pi$ are summarized in Table~\ref{noetac_sys}.

\begin{table}[tb]
	\centering
	\caption{Relative systematic uncertainties (\%) in the measurement of the product branching fraction $\mathcal{B}(\psi(2S)\rightarrow \gamma \eta_{c}(2S))\times\mathcal{B}(\eta_{c}(2S)\rightarrow \pi^{+}\pi^{-}K^{0}_{S}K\pi) $.}
	\label{noetac_sys}
	\begin{tabular}{l|c}
		\hline
		\multicolumn{1}{l|}{Source}         & Uncertainty  \\ \hline \hline
		$N_{\psi(2S)}$                      & 0.5          \\ \hline 
		Tracking efficiency                 & 6.0          \\ \hline
		Photon reconstruction               & 1.0          \\ \hline
		$K^{0}_{S}$ reconstruction          & 1.0          \\ \hline
	    Kinematic fit                       & 2.0          \\ \hline
		$J/\psi$ veto                       & 3.8          \\ \hline
    	Fit range                           & 3.8          \\ \hline
		Signal shape                        & 18.9         \\ \hline
		Background estimation               & 21.5         \\ \hline
				
		Total                               & 29.8         \\ \hline
		
	\end{tabular}
\end{table}

  The systematic uncertainty due to the tracking efficiency, photon reconstruction, $K_{S}^0$ reconstruction, total number of $\psi$(2S) events, kinematic fit, $J/\psi$ veto, fit range, and signal shape are estimated by the same method introduced in Sec.~\ref{sysA}. 

The systematic uncertainty due to the background is estimated by changing the relative ratio $f_{FSR}$  and changing the  parameterisation of continuum background.


\section{Summary}

Based on (27.12± 0.14)$\times 10^8$ $\psi$(2S) events collected by the BESIII detector, we search for the decays $\eta_{c}(2S) \rightarrow \pi^{+}\pi^{-}K^{0}_{S}K^{\pm}\pi^{\mp}$, and $\eta_c (2S)\rightarrow \pi^{+}\pi^{-}\eta_c$  with $\eta_c\rightarrow K_S^0 K^{\pm}\pi^{\mp}$ and $\eta_c\rightarrow K^{+}K^{-}\pi^{0}$.
 No significant signal of $ \eta_{c}$(2S) $\rightarrow \pi^{+}\pi^{-}\eta_c$ is observed, and the upper limit of the product branching fraction $\mathcal{B}$($\psi(2S)\rightarrow \gamma \eta_{c}(2S))\times \mathcal{B}(\eta_c (2S) \rightarrow \pi^{+} \pi^{-} \eta_c$) is determined to be $2.21\times10^{-5}$ at the 90\% C.L.. 
The $\eta_{c}(2S) \rightarrow \pi^{+}\pi^{-}K^{0}_{S}K^{\pm}\pi^{\mp}$ decay is observed with a statistical significance of 10$\sigma$ for the first time. The $\mathcal{B}(\eta_{c}(2S)\rightarrow \pi^{+}\pi^{-}K^{0}_{S}K^{\pm}\pi^{\mp})$ is measured to be ($1.33 \pm 0.11\pm 0.4\pm 0.95$)$\times 10^{-2}$, where the first and
second uncertainties are statistical and systematic, respectively, and the third uncertainty is from the quoted $\mathcal{B}(\psi(2S)\rightarrow \gamma \eta_{c}(2S))$~\cite{pdg}. The product branching fraction $\mathcal{B}(\psi(2S)\rightarrow \gamma \eta_{c}(2S))\times\mathcal{B}(\eta_{c}(2S)\rightarrow \pi^{+}\pi^{-}K^{0}_{S}K\pi) $ is determined to be $(9.31 \pm 0.72\pm 2.77)\times 10^{-6}$, which is consistent with the previous measurement~\cite{yliang}.

\section*{\boldmath ACKNOWLEDGMENTS}

The BESIII Collaboration thanks the staff of BEPCII and the IHEP computing center for their strong support. This work is supported in part by National Key R\&D Program of China under Contracts Nos. 2023YFA1606704, 2020YFA0406300, 2020YFA0406400; National Natural Science Foundation of China (NSFC) under Contracts Nos. 11635010, 11735014, 11835012, 11935015, 11935016, 11935018, 11961141012, 12025502, 12035009, 12035013, 12061131003, 12192260, 12192261, 12192262, 12192263, 12192264, 12192265, 12221005, 12225509, 12235017; the Chinese Academy of Sciences (CAS) Large-Scale Scientific Facility Program; the CAS Center for Excellence in Particle Physics (CCEPP); Joint Large-Scale Scientific Facility Funds of the NSFC and CAS under Contract No. U1832207; CAS Key Research Program of Frontier Sciences under Contracts Nos. QYZDJ-SSW-SLH003, QYZDJ-SSW-SLH040; 100 Talents Program of CAS; The Institute of Nuclear and Particle Physics (INPAC) and Shanghai Key Laboratory for Particle Physics and Cosmology; European Union's Horizon 2020 research and innovation programme under Marie Sklodowska-Curie grant agreement under Contract No. 894790; German Research Foundation DFG under Contracts Nos. 455635585, Collaborative Research Center CRC 1044, FOR5327, GRK 2149; Istituto Nazionale di Fisica Nucleare, Italy; Ministry of Development of Turkey under Contract No. DPT2006K-120470; National Research Foundation of Korea under Contract No. NRF-2022R1A2C1092335; National Science and Technology fund of Mongolia; National Science Research and Innovation Fund (NSRF) via the Program Management Unit for Human Resources \& Institutional Development, Research and Innovation of Thailand under Contract No. B16F640076; Polish National Science Centre under Contract No. 2019/35/O/ST2/02907; The Swedish Research Council; U. S. Department of Energy under Contract No. DE-FG02-05ER41374


\clearpage
\bibliography{main}

\end{document}